\documentclass[fleqn,usenatbib,useAMS]{mnras}

\usepackage{newtxtext,newtxmath}

\usepackage[T1]{fontenc}
\usepackage{ae,aecompl}

\usepackage{graphicx}	

\usepackage{comment}
\usepackage{color}

\title [Synthesis of infrared Stokes spectra in an evolving chromospheric jet]{Synthesis of infrared Stokes spectra in an evolving solar chromospheric jet}

\author[T. Matsumoto et al.]{
T. Matsumoto,$^{1,2}$\thanks{E-mail: takuma.matsumoto@isee.nagoya-u.ac.jp}
Y. Kawabata,$^{2}$
Y. Katsukawa$^{2}$
H. Iijima$^{1}$
and C. Quintero Noda$^{3,4}$
\\
$^{1}$Centre for Integrated Data Science, Institute for Space-Earth Environmental Research, Nagoya University, Furocho, Chikusa-ku, Nagoya, Aichi 464-8601, Japan\\
$^{2}$National Astronomical Observatory of Japan, 2-21-1 Osawa, Mitaka, Tokyo 181-8588, Japan\\
$^{3}$Instituto de Astrof\'isica de Canarias, E-38200, La Laguna, Tenerife, Spain \\
$^{4}$Departamento de Astrof\'isica, Univ. de La Laguna, La Laguna, Tenerife, E-38205, Spain}

\date{Accepted XXX. Received YYY; in original form ZZZ}

\pubyear{2015}

\begin{document}
\label{firstpage}
\pagerange{\pageref{firstpage}--\pageref{lastpage}}
\maketitle

\begin{abstract}
Chromospheric jets are plausible agents of energy and mass transport in the solar chromosphere, although their driving mechanisms have not yet been elucidated.
Magnetic field measurements are key for distinguishing the driving mechanisms of chromospheric jets.
We performed a full Stokes synthesis in the infrared range with a realistic radiative magnetohydrodynamics simulation that generated a chromospheric jet to predict
spectro-polarimetric observations from the Sunrise Chromospheric Infrared spectro-Polarimeter (SCIP) onboard the SUNRISE III balloon telescope.
The jet was launched by the collision between the transition region and an upflow driven by the ascending motion of the twisted magnetic field at the envelope of the flux tube.
This motion is consistent with upwardly propagating non-linear Alfv\'{e}nic waves.
The upflow could be detected as continuous Doppler signals in the \ion{Ca}{ii} 849.8 nm line at the envelope where the dark line core intensity and strong linear polarisation coexist.  
The axis of the flux tube was bright in both \ion{Fe}{i} 846.8 nm and \ion{Ca}{ii} 849.8 nm lines with down- flowing plasma inside it.
The structure, time evolution, and Stokes signals predicted in our study will improve the physical interpretation of future spectro-polarimetric observations with SUNRISE III/SCIP.
\end{abstract}


\section{Introduction }

Chromospheric jets are collimated eruptions that exist ubiquitously in the chromosphere.Various types of jets, such as spicules \citep{1968SoPh....3..367B}, surges \citep{1973SoPh...28...95R}, and anemone jets \citep{2007Sci...318.1591S} have been reported so far.
Significant amounts of energy and mass are considered to be transported via chromospheric jets,
however, the driving mechanisms have not yet been elucidated.
Because the magnetic field plays a crucial role in dynamics, 
spectro-polarimetry could be the key observation to determine the driving mechanism.

The collision between the magnetohydrodynamics (MHD) shock and transition region is a promising mechanism for accelerating chromospheric jets \citep{1961ApJ...134..347O}.
When a certain amount of energy is released in the deep layer, it produces an acoustic wave that propagates upward.
As the acoustic wave propagates upward, it evolves into a shock owing to the atmospheric stratification and hits the transition region \citep{1982SoPh...77..121S}.
When the shock hits the transition region, it breaks into shocks and a contact discontinuity, below which the chromospheric materials are lifted.

There are several candidates for the origin of the MHD shock.
First, the magnetic pressure and centrifugal force associated with Alfv\'{e}n waves generate magneto-acoustic waves via nonlinear mode conversion \citep{1982SoPh...75...35H,1999ApJ...514..493K}.
Horizontal magnetic signals followed by propagating shocks are expected in this process.
Second, sudden downflow is caused by processes, such as magnetic pumping \citep{2011ApJ...730L..24K,2016ApJ...827....7K} or convective collapse \citep{1978ApJ...221..368P,1979SoPh...61..363S} that pull down the plasma in the flux tubes to produce rebounding slow shocks.
When the flux tubes are twisted owing to a photospheric vortex \citep{1988Natur.335..238B,2008ApJ...687L.131B}, the perturbed twist propagates with slow shock.
In such a case, it might be difficult to distinguish between the Alfv\'{e}n wave model and the rebound shock model because both processes may contribute to feeding the magneto-acoustic waves.
Third, magnetic reconnection between pre-existing  and newly emerged fields will be the agent for energy release in the chromosphere \citep{2007Sci...318.1591S}.
The photospheric magnetic features, as well as the chromospheric current structure, could provide key evidence for reconnection.

To determine the origin of chromospheric jets, we need to infer the properties of the magnetised atmosphere at multiple layers, which requires observing spectral lines that form at different heights. Unfortunately, these transitions also fall far apart in the wavelength spectrum, strictly simultaneously requiring the observation of the Sun at multiple wavelengths.
Polarimetric observations will provide magnetic field information, such as mixed polarity at the foot point, current concentrations near the magnetic inversion lines, and helical field, including rich information in the driving process.
Line-of-sight structures also help in understanding the height of the energy source that can be deduced from the localised temperature increase \citep{2018A&A...609A..14R} and bidirectional flow \citep{2019ApJ...883..115N}, and propagating wave signals \citep{2020A&A...643A.166T}.
The dynamical nature of polarimetric signals in the chromosphere has been revealed in several contexts \citep{2018A&A...619A..63J,2020A&A...642A.128S,2022A&A...664A...8M}; the demands for theoretical predictions for polarimetric signals in the chromospheric jets are increasing to properly interpret the complex observations.
The theoretical predictions will help interpret the upcoming polarimetric observations with
a wide spectral range, as well as high polarimetric sensitivity, such as SUNRISE III\footnote{Official website of the SUNRISE III project can be found in  https://www.mps.mpg.de/solar-physics/sunrise}, Daniel K. Inouye Solar Telescope \citep[DKIST]{2020SoPh..295..172R}, and European Solar Telescope \citep[EST]{2022A&A...666A..21Q}, which provides smoking-gun evidence for the origin of the chromospheric jets.

This study aims to predict the Stokes signals that will be observed by SUNRISE III/SCIP \citep[Sunrise Chromospheric Infrared spectro-Polarimeter]{2020SPIE11447E..0YK}.
SCIP will be deployed aboard the third flight of SUNRISE stratospheric balloon-borne solar telescope \citep{2011SoPh..268....1B}.
The instrument is a slit scanning spectro-polarimeter capable of simultaneously observing two wide spectral bands, covering 770 nm and 850 nm bands.
These bands include both photospheric lines, such as \ion{Fe}{i} (846.8 nm) and \ion{K}{i} (766.5 nm \& 769.9 nm), as well as lower chromospheric lines, such as \ion{Ca}{ii} (849.8 nm \& 854.2 nm).
Additionally, SCIP features a high degree of polarization accuracy of $3\times10^{-4}$ (1 sigma) and a high spatial resolution of 0.2 arcsec, owing to negligible effect of atmospheric seeing, providing a powerful tool for studying the fine-scale magnetic fields in the solar chromosphere.
By utilizing the unique capabilities of SCIP, this study has the potential to deepen our understanding of the magnetic activities in the solar chromosphere.

Forward modeling using full Stokes synthesis of radiative MHD simulations is a robust tool for interpreting complex observed spectra and simulating the capabilities of a given instrument or mission.
A series of studies have been devoted to predicting the potential of the spectral lines to be observed by SCIP to infer the properties of, e.g., magnetic pumping \citep{2017MNRAS.472..727Q} and chromospheric jets \citep{2019MNRAS.486.4203Q}.
Although polarisation signals from the helical magnetic field associated with chromospheric jets are expected to be detectable with SUNRISE III/SCIP, their origin and relationship with the jet have not been investigated yet.
Hence we expand on the previous work using a time-series analysis to investigate further the driving processes and the evolution of the chromospheric jet.

Our analysis suggests that Alfv\'{e}nic motion in the flux tube generates an upflow that drives the chromospheric jet. 
Consequently, strong linear polarisation with dark, blue-shifted, and arc-like structures around the flux-tube axis will be detectable with \ion{Ca}{ii} infrared line.
Along the axis of the flux tube, bright and red-shifted cores are continuously observed in both \ion{Fe}{i} (846.8 nm) and \ion{Ca}{ii} (849.8 nm) lines.
Careful consideration of observation configurations, including the field of views and integration time in SUNRISE III/SCIP can capture all these features and help distinguish the origin of the jets.

This paper is organised as follows:
Section 2 presents methods and dataset, and 
Section 3 details our data analysis and results with discussions.
Finally, the conclusions of the study can be found in Section 4.

\section{Methods and Dataset}
\begin{figure}
 \includegraphics[width=\columnwidth]{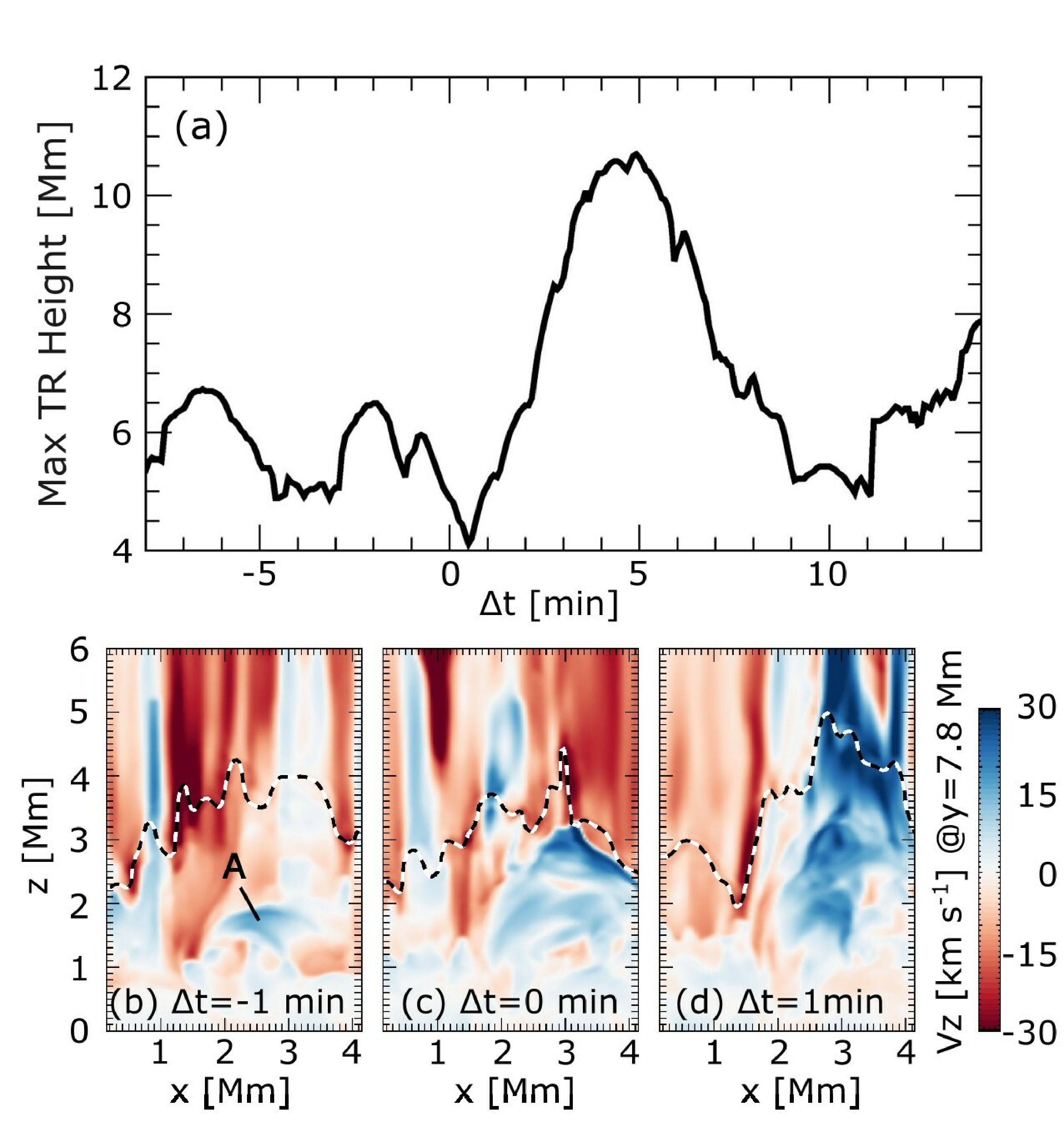}
 \caption{Summary of the time evolution of the analysed jet. (a) Height of the transition region whose temperature was defined to be 4$\times10^4$ K (The maximum value was taken in x = 2.7 Mm plane) as a function of time. (b)--(d) vertical velocity, $v_{\rm z}$ at y = 7.8 Mm. The dashed lines in (b)--(d) indicate the height of the transition region. Temporal evolution ($\Delta$t = -1, 0, 1 min) is displayed from left to right.
 Blue represents upflows while red colour indicates downflowing materials.}
 \label{fig:mhd_evolution_2d}
\end{figure}

In this study, the full Stokes profiles were derived using a radiative MHD simulation combined with a Stokes synthesis code.
Infrared spectra emerging from a simulated chromospheric jet were modeled to predict future observations from SUNRISE III/SCIP.
Time-series analysis of the synthesised spectra enabled us to distinguish the driving mechanisms of the jet that could not be clarified by spectropolarimetry with a single snapshot.

\subsection{Numerical simulation}
As the reference solar atmosphere, we used a dataset from a radiative MHD simulation identical to that used in \cite{2017ApJ...848...38I}.
The simulation was conducted using RAMENS code, and the full details of the simulation can be found in \citep{2015ApJ...812L..30I,2016PhDT.........5I}.
In brief, the RAMENS code is dedicated to simulating a realistic solar atmosphere in which various types of physics are coupled.
In addition to the ideal MHD equations, gravity, Spitzer-type thermal conduction, equation of state under the assumption of LTE, including the latent heat of partial ionisation, and radiative cooling are included in the model.
Radiative cooling is computed as a combination of optically thick radiative cooling with gray approximation and optically thin cooling.
Based on these assumptions, the RAMENS code enables us to provide a realistic solar atmosphere.

The MHD dataset represents a unipolar quiet region with an average magnetic flux density of 10 G.
The total duration analysed was approximately 22 min (t $\in$ [396, 418] min) with a temporal cadence of 5 s.
The horizontal domain is 9 Mm for in both the x- and y-directions with a periodic boundary.
The vertical domain ranges from -2 to 14 Mm and includes both the surface convection zone and corona. 
The height z = 0 Mm roughly corresponds to the surface where the optical depth is unity at the continuum wavelengths at 500 nm.

Among the chromospheric jets appearing in the dataset, we focused on the jet investigated in \cite{2017ApJ...848...38I}.
The jet started to erupt from t = 404 min to t = 413 min, showing a parabolic trajectory (Fig. \ref{fig:mhd_evolution_2d}a).
Hereafter, we use a different time coordinate, $\Delta {\rm t}~\equiv {\rm t}-404$ min, to set $\Delta $t = 0 when the jet started for convenience.
Starting from $\Delta$t = 0, the jet reached its peak height of approximately 11 Mm at $\Delta$t = 5 min and ended at $\Delta$t = 9 min.
A photospheric magnetic sheet existed below the jet to form a flux tube before the formation of the jet.
At $\Delta$t = -1 min, there appears to be a blob of upflow (the upflow A in Fig. \ref{fig:mhd_evolution_2d}b) that will collide with the transition region to drive the jet at $\Delta$t = 0 min (Fig. \ref{fig:mhd_evolution_2d}c).
After upflow A hits the transition region, chromospheric materials are lifted to form the jet.

\subsection{Synthesis of the Stokes profiles}

We synthesised full-Stokes profiles in the infrared range, including the \ion{Ca}{ii} 849.8 nm and \ion{Fe}{i} 846.8 nm lines.
We focused on the \ion{Ca}{ii} 849.8 nm line rather than the more commonly used 854.2 nm line in this study because the polarisation signals turned out to be stronger in the 849.8 nm line, although SCIP can observe both spectral lines simultaneously.
The RH code was used to synthesise these lines \citep{2001ApJ...557..389U}.
The equations of radiative transfer and statistical equilibrium under non-LTE conditions are solved in the RH code.
Among the four modules specific for different geometries, rhf1d was used to calculate the emergent spectra assuming a 1D plane-parallel geometry.
This geometry module allows us to calculate the Stokes profiles on a column-by-column basis, which can be appropriate for the \ion{Ca}{ii} lines where horizontal scattering does not have significant effects \citep{2009ApJ...694L.128L}.
We used the RH code version 2, last modified on 2020 May 1.

Several modifications have been made to the input files that are necessary for the RH code.
First, complete redistribution was assumed for all spectral lines.
Second, the LTE hydrogen population was calculated using an abundance identical to that used in the RAMENS code.
The rest of configurable parameters were identical to their default settings.

A spatial and spectral degradation was applied to  the synthetic Stokes profiles to reproduce the spatial and spectral resolution of SUNRISE III/SCIP.
For the spatial and spectral convolution, a Gaussian kernel with an FWHM of 0.2 $\times~(\lambda/854~{\rm nm})$ arcsec and 4 pm were applied, respectively. 
Although the spectral resolution has a small dependence on wavelength, we used a constant value of 4 pm for simplicity.
Because there were several pixels with poor convergence in Stokes synthesis (less than 1\%), we ignored these pixels during the convolution process.

To simplify the analysis, six representative variables were mainly used rather than showing all the Stokes profiles: the intensity at the line core at the rest wavelength, $I_{\rm core}$, Doppler velocity $V_{\rm LOS}$, maximum amplitude of circular (MCP) and linear (MLP) polarisation, total linear polarisation (TLP), and linear polarisation azimuth (LPA).
The centre-of-mass wavelength of $I_{\rm cont} - I$ at 11 spectral points near the line centre is used to determine the line-of-sight velocity for each line.
MCP and MLP are defined as follows:
\begin{eqnarray}
 \mathrm{MCP} &\equiv& \underset{|\lambda-\lambda_0| < \Delta \lambda}{\rm max} (|V(\lambda)|), \\
 \mathrm{MLP} &\equiv& \underset{|\lambda-\lambda_0| < \Delta \lambda}{\rm max} \left( \sqrt{Q(\lambda)^2 + U(\lambda)^2}\right) 
\end{eqnarray}
where $\Delta \lambda $ = 20 pm (five spectral sampling sizes) for each line and $\lambda_0$ is the central wavelengths in the rest frame.
The TLP is defined as 
\begin{eqnarray}
  {\rm TLP} \equiv \sqrt{ \langle Q \rangle ^2 + \langle U \rangle ^2}, 
\end{eqnarray}
where 
\begin{eqnarray}
 \langle f \rangle \equiv \frac{1}{2\Delta \lambda} \int ^{\lambda_0 + \Delta \lambda} _{\lambda_0 - \Delta \lambda} f (\lambda) d\lambda.
\end{eqnarray}
Following the equations given in \cite{2004ASSL..307.....L}, LPA, which provides an estimation of the magnetic field azimuth, is defined as
\begin{eqnarray}
 \phi = \frac{1}{2}\arctan \left( \frac{U}{Q}\right),
\end{eqnarray}
where spectral points with maximum amplitudes of $Q$ and $U$ in $|\lambda-\lambda_0|<\Delta \lambda$ were used.

One of the main advantages of our approach is that it analyses the time series of Stokes profiles throughout the entire evolution of the jet.
With this approach, we attempted to specify the spectral properties from the driving process of a jet that were not obtained in the previous study of \cite{2019MNRAS.486.4203Q}.

\section{Results and Discussion} \label{sec:results} 
Overall, the results presented below show that the jet was driven by a blob of upflow connected to the twisted field lines above a flux tube. 
A bright down-flowing core and dark up-flowing envelope structure with strong linear polarisation signals can be detected using spectro-polalimetric observations from SUNRISE III/SCIP.

\begin{figure}
 \includegraphics[width=\linewidth]{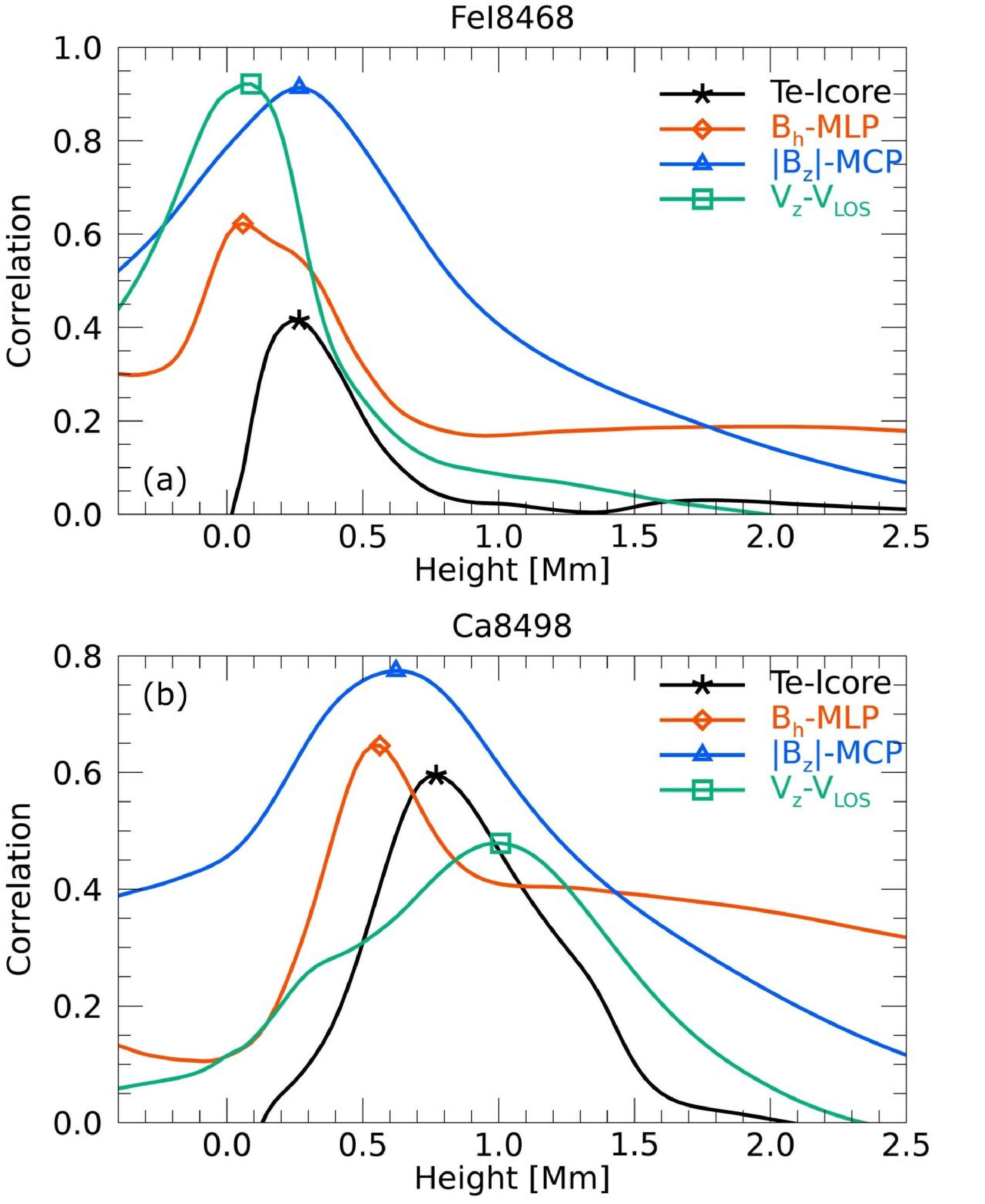}
 \caption{Pearson's product moment correlations between synthetic data in (a) \ion{Fe}{i} and (b) \ion{Ca}{ii} line and MHD variables as a function of height.
 The black, orange, blue, and green lines represent the correlation between temperature \& $I_{\rm core}$, horizontal field strength ($B_{\rm h} \equiv \sqrt{B_{\rm x}^2 + B_{\rm y}^2}$) \& MLP, vertical field strength \& MCP, and vertical velocity \& $V_{\rm LOS}$, respectively.
 Symbols designate the height where each correlation is maximum.}
 \label{fig:formation_height}
\end{figure}
\subsection{Estimation of formation layers}
To roughly estimate the formation layers of each line, Pearson's product moment correlations were obtained between temperature \& $I_{\rm core}$, horizontal field strength \& MLP, vertical field strength \& MCP, and vertical velocity \& $V_{\rm LOS}$ at each height.
The peaks in the correlations from \ion{Fe}{i} line ranged in 0--0.3 Mm (Fig. \ref{fig:formation_height}a), while those from the \ion{Ca}{ii} line ranged in 0.5--1.0 Mm (Fig. \ref{fig:formation_height}b).
These analyses will help us understand the relationship between vertical structures and synthetic data, although further analysis is necessary to retrieve exact height information

\subsection{Evolution of the synthesized observables during the jet}
\begin{figure*}
 \includegraphics[width=180 mm]{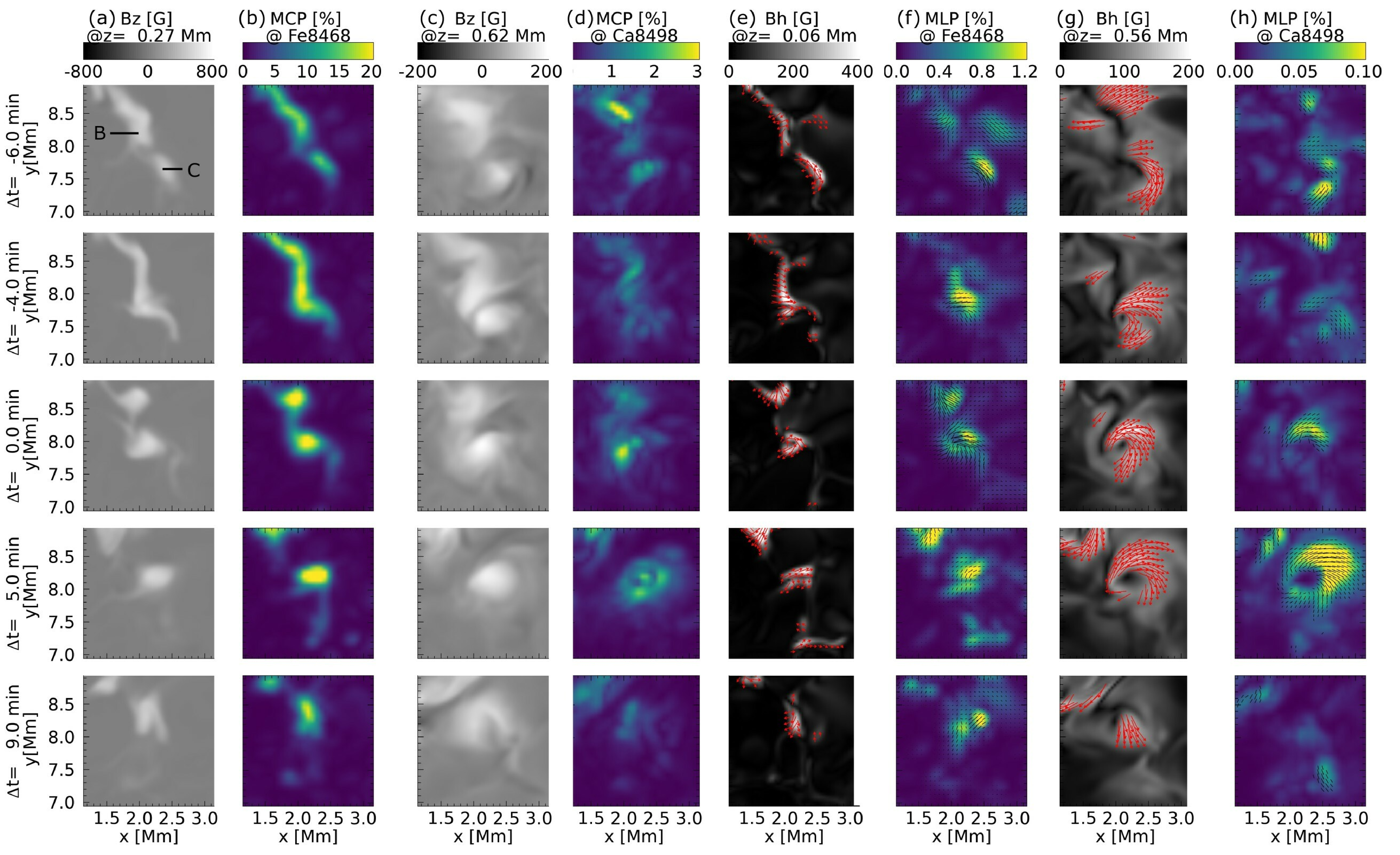}
 \caption{Time evolution of the synthetic images (MCP \& MLP) and the related MHD variables during the jet (- 6 min $<$ $\Delta$t $<$ + 9 min). From left to right column, (a) $B_{\rm z}$ at z = 0.27 Mm, (b) MCP for the \ion{Fe}{i} line, (c) $B_{\rm z}$ at z = 0.62 Mm, (d) MCP for the \ion{Ca}{ii} line, (e) $B_{\rm h}$ at z = 0.06 Mm, (f) MLP for the \ion{Fe}{i} line, (g) $B_{\rm h}$ at z = 0.56 Mm, and (h) MLP for the \ion{Ca}{ii} line were shown. The red arrows in columns (e) and (g) indicate the horizontal component of magnetic field vector. The black line segments in columns (f) and (h) represent the LPA where polarization signals were larger than 3$\times$10$^{-4}$ of $I_{\rm cont}$.}
 \label{fig:xy_synthesis_map}
\end{figure*}

The synthetic MCP signals capture the evolution of the axial field of the jet (columns a--d in Fig. \ref{fig:xy_synthesis_map}).
From $\Delta {\rm t} = - 6 {\rm ~min}$, the two flux sheets (B \& C in column a in \ref{fig:xy_synthesis_map}) in the granular lanes approached the intersection of the granular lanes.
The lower half of sheet B merged with sheet C at $\Delta$t = - 4 min to create a flux tube where the foot point of the jet is located.
The upper half of sheet B was moved up to create another flux tube.
These flux sheets produced strong polarisation signals well above 10$^{-3}$ of $I_{\rm cont}$ on \ion{Fe}{i} line during the lifetime of the jet.
A cadence of 1 min is sufficient to capture the flux-sheet merger ($\Delta$t = - 4 min).
At a height of z = 0.6 Mm, flux sheet B expanded to create a round-shaped axial field of the jet.
The axial field was surrounded by a negative spiral-like structure after the flux-sheet merger (column c in Fig. \ref{fig:xy_synthesis_map}).
We found that the negative structure was a part of the twisted field, whose elevation angle was locally negative.
Although the axial-field components have strong polarisation signals in the \ion{Ca}{ii} line, 
a twisted field with a negative LOS component exhibited weak signals (column d in Fig. \ref{fig:xy_synthesis_map}).

The twisted field structure around the axis of the jet was spatially resolved with reasonable linear polarisation signals (columns e--h in Fig. \ref{fig:xy_synthesis_map}) that are above the noise level expected for SCIP.
At the photospheric height, the twisted field around the flux tube appears after the flux merger ($\Delta$t = - 4 min) and gradually 
disappears after the jet reaches its maximum height ($\Delta$t = 5 min).
The photospheric twist revealed strong linear polarisation signals around the envelope of the axial field.
A tornado-like structure is continuously observed at the chromospheric heights (column g in Fig. \ref{fig:xy_synthesis_map}).
The corresponding structure was found in MLP signals, although the twist direction was not available owing to the 180° ambiguity.
The magnetic twist became increasingly prominent until $\Delta$t = 5 min, and gradually disappeared by $\Delta$t = 9 min (column h in Fig. \ref{fig:xy_synthesis_map}).

\begin{figure*}
 \includegraphics[width=180 mm]{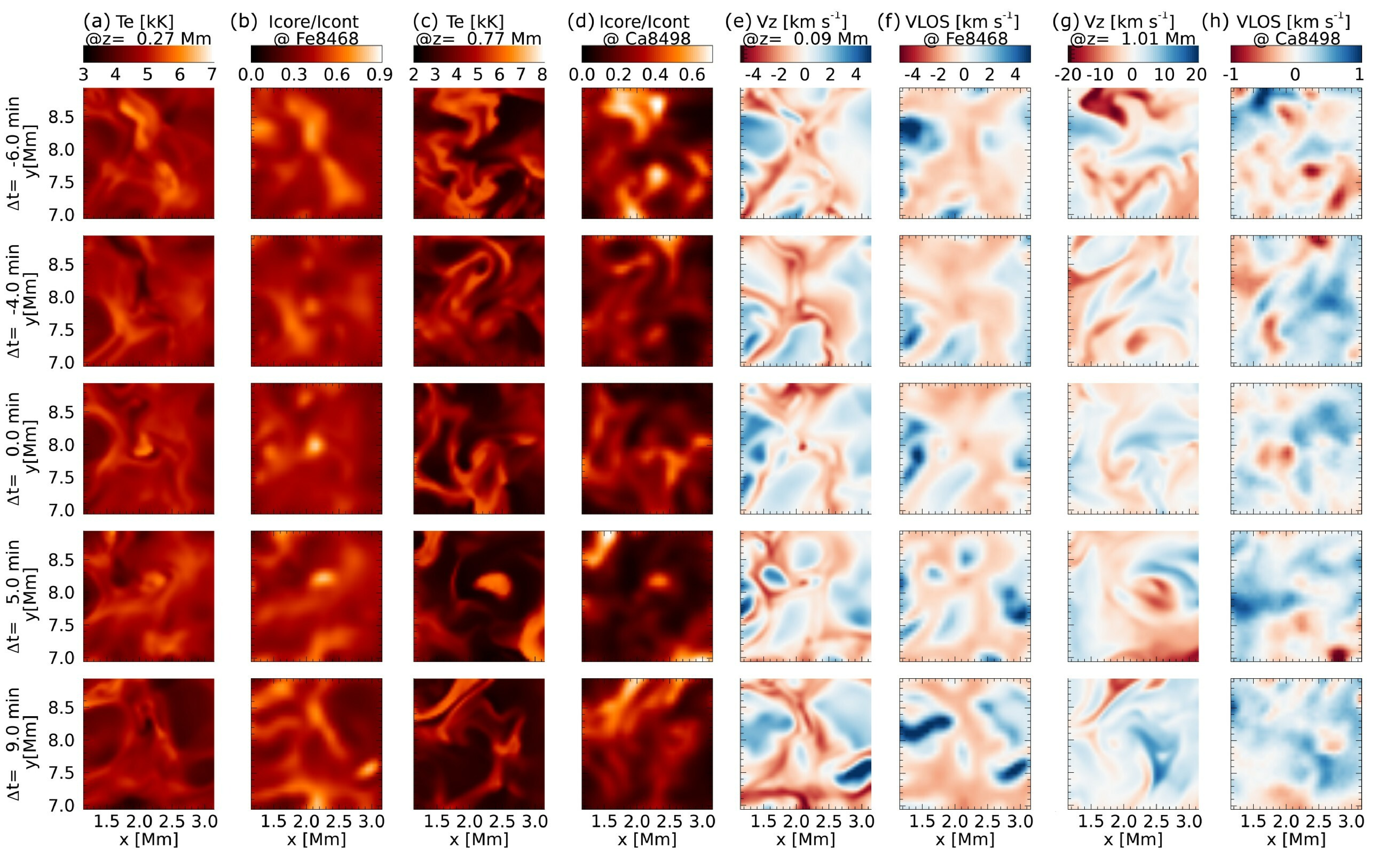}
 \caption{Time evolution of the synthetic images ($I_{\rm core}$ \& $V_{\rm LOS}$ ) and the related MHD variables during the jet (-6 min $<\Delta$t$<$ + 9 min). From left to right column, (a) Te at z = 0.27 Mm, (b) $I_{\rm core}$ for the \ion{Fe}{i} line, (c) Te at z = 0.77 Mm, (d) $I_{\rm core}$ for the \ion{Ca}{ii} line, (e) $V_{\rm z}$ at z = 0.09 Mm, (f) $V_{\rm LOS}$ for the \ion{Fe}{i} line, (g) $V_{\rm z}$ at z = 1.01 Mm, and (h) $V_{\rm LOS}$ for the \ion{Ca}{ii} line were shown. }
 \label{fig:xy_synthesis_map2}
\end{figure*}

From the line-core images, we can infer that the jet had a hot core and cool envelopes (columns a--d in Fig. \ref{fig:xy_synthesis_map2}).
A bright spot of approximately 0.2 Mm appeared in the line core intensity of \ion{Fe}{i} line above the flux tube axis after the flux merger.
This enhancement continued until at least $\Delta$t = 5 min, and disappeared at $\Delta$t = 9 min.
The temperature at the photospheric heights also exhibited similar behaviour.
At the chromospheric heights, hot core and cool envelope appeared well before and after the jet initiation ($\Delta$t = -6,-4, and 5 min).
The corresponding structure was found in the line-core images of \ion{Ca}{ii} as a bright core and dark envelope.
This feature sometimes can be observed as magnetic tornadoes or chromospheric swirls found in past observations \citep{2009A&A...507L...9W,2012Natur.486..505W}.

The Doppler signals help to find the descending core and ascending envelope of the jet (columns e--h in Fig. \ref{fig:xy_synthesis_map2}).
The granulation pattern was observed in the Doppler signals of the \ion{Fe}{i} line.
There was a small downflow near the axis of the flux tube, particularly after the flux sheet merger, which is consistent with the magnetic vortex appearing in the previous simulation \citep{2013ApJ...770...37K}.
At the chromospheric heights, the flux tube axis tended to have a downward flow, while the envelope tended to have an upflow in the \ion{Ca}{ii} line.
However, it was challenging to distinguish the upflow A in the x-y plane at the chromospheric heights from a single snapshot.
This was because the fluctuations in the vertical velocity were dominated by acoustic waves that hindered the determination of the velocity in the jet.
A blue shift was sometimes obtained at the same location as the jet-related upflow, although a time-series analysis, such as sit and stare observation, was necessary to confirm that the blue shift was related to the jet-upflow.

\subsection{Sit-and-stare observation}
\begin{figure}
 \includegraphics[width=\columnwidth]{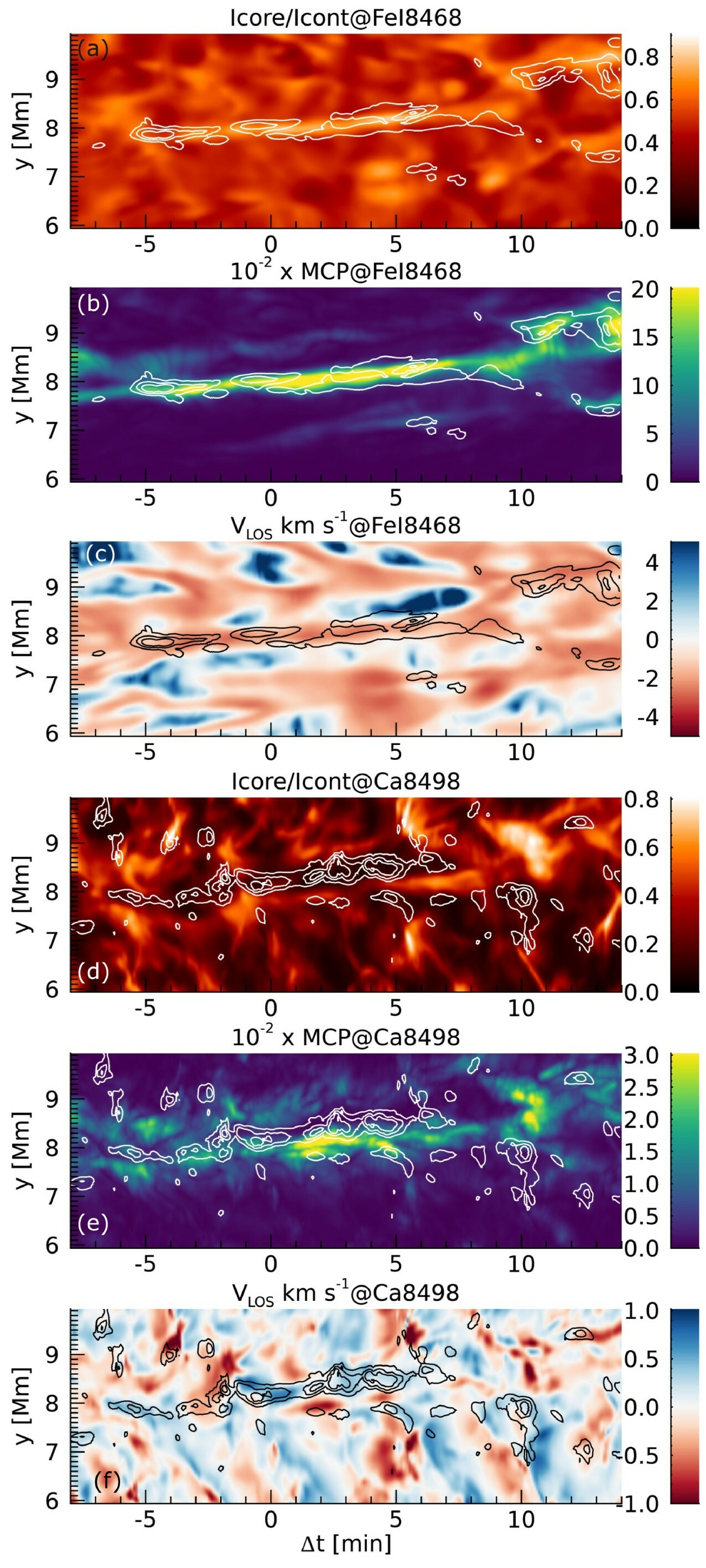}
 \caption{Time-distance diagram of (a) $I_{\rm core}$, (b) MCP, and (c) $V_{\rm LOS}$ derived from the synthetic spectra of the \ion{Fe}{i} 846.8 nm line, and (d) $I_{\rm core}$, (e) MCP, and (f) $V_{\rm LOS}$ derived from the synthetic spectra of the \ion{Ca}{ii} 849.8 nm line. Contours of 2, 4, and 6 (1, 2, and 3) sigma levels in MLP were over-plotted in the panels from (a) to (c) ( (d) to (f) ) . The slit position was taken along x = 2.1 Mm.}
 \label{fig:sit_and_stare}
\end{figure}

A set of time distance diagrams for the \ion{Fe}{i} 846.8 nm and \ion{Ca}{ii} 849.8 nm lines were derived to simulate a sit-and-stare observation (Fig. \ref{fig:sit_and_stare}).
The slit was located at the centre of the photospheric magnetic concentration (x = 2.1 Mm) along the y-direction.
At the photospheric heights, we found that the localised enhancement of the line core intensity continued throughout the lifetime of the jet (Fig. \ref{fig:sit_and_stare}a).
At the same location, the continuous MCP signals overlapped with the line-core enhancement with strong MLP signals at the envelope (Fig. \ref{fig:sit_and_stare}b).
Moreover, the down-flow signals with 2--3 km s$^{-1}$ were associated with the hot and magnetised core of the flux tube (Fig \ref{fig:sit_and_stare}c).
At the chromospheric heights, 
both the dark envelope and bright core can be well distinguished, especially in the later phase of the jet ($\Delta$t > 5 min) (Fig \ref{fig:sit_and_stare}d).
The axial field was detectable using MCP because the signal of approximately 2 \% of $I_{\rm cont}$ was well above the polarimetric sensitivity of SUNRISE III/SCIP ($1\sigma = 3\times 10^{-4}$ of $I_{\rm cont}$) (Fig. \ref{fig:sit_and_stare}e).
The amplitude of MLP signals in the arc-like structure was weaker, so it would be
difficult to distinguish between noise and signals using MLP signals for this case.
Because the arc-like structure had a width of 0.5 Mm and a lifetime of more than 5 min, spatio-temporal binning will increase the S/N ratio.
The arc structure in MLP originated from the field lines at the envelope of the flux tube is writhing counterclockwise.
An upward Doppler velocity of approximately 1 km s$^{-1}$ above the arc indicates the upward motion of the writing flux tube (Fig. \ref{fig:sit_and_stare}f). 
Because the fluctuations of the Doppler velocity in the ambient region were comparable to those of the flux tube, it was necessary to combine the Doppler and polarimetric signals to infer the velocity of the flux tube.

\subsection{Detection of linear polarization signals}
\begin{figure}
 \includegraphics[width=\columnwidth]{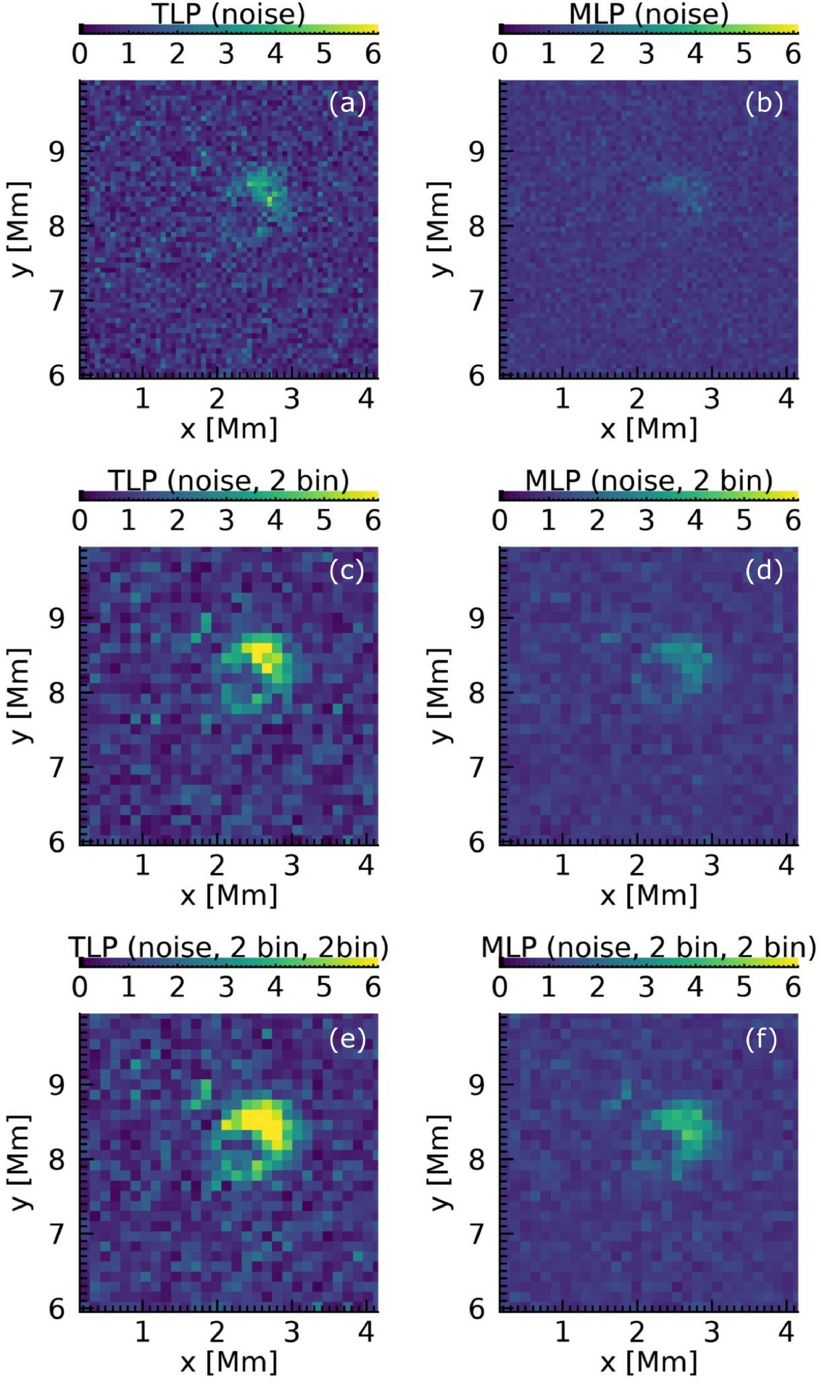}
 \caption{Linear polarization signals from the \ion{Ca}{ii} 849.8 nm line at $\Delta$t = 5 min: (a) TLP with noise, (b) MLP with noise, (c) TLP with noise, 2 binned in the spatial domain, (d) MLP with noise, 2 binned in spatial domain (e) TLP with noise, 2 binned both in the spatial and time domain, (f) MLP with noise, 2 binned in the spatial and time domain. The colour corresponds to S/N ratio. For temporal binning, data in $\Delta$t = 5 min - 10 sec was also used.}
 \label{fig:degraded}
\end{figure}

Analysing the linear polarisation signals in chromospheric lines is challenging because they are generally less sensitive to the magnetic field due to their lower effective Land\'{e} factors.
So, in our case, it may be advisable to use the reference variable TLP instead of MLP to analyse the \ion{Ca}{ii} linear polarisation signals.
This is because the averaging operation over wavelength reduces the noise level, whereas the maximum operation increases the noise level.
For instance, when applying Gaussian noise with 4 $\times$ 10$^{-4}$ I$_{\rm cont}$, both MLP and TLP signals became comparable to the noise level (Fig. \ref{fig:degraded}a \& b).
Performing a spatial binning (the spatial sampling increasing from 0.1 to 0.2 arcsec) increased the signal-to-noise ratio up to 6 (Fig. \ref{fig:degraded}c \& d).
The arc structure was still visible owing to the spatial scale of approximately 1 Mm, although further binning destroyed the structure.
Performing an additional binning on the time domain ( temporal sampling increases from 10 to 20 s) further increases the signal-to-noise ratio to more than 6 (Fig. \ref{fig:degraded}e \& f) in the most prominent phase.

To optimize the observations with SCIP, it is important to consider the trade-off between temporal cadence and polarimetric accuracy, given that SCIP is a slit scanning spectro-polarimeter. 
In this study, the magnetic arc had a spatial scale of approximately 1 Mm, and its evolution could be effectively resolved with a cadence of less than 2 minutes. 
By selecting an integration time of 10.24 seconds, the slit scan cover an area of 1 arcsec times the slit size in 1.9 minutes.
In addition, that integration times will produce a polarization accuracy of $4\times10^{-4}$ of $I_{\rm cont}$ (1 sigma) without spatial nor temporal summing in the \ion{Ca}{ii} 849.8 nm line. 
Although the FOV is narrower than the magnetic arc, a S/N ratio greater than 6 can be obtained by spatially summing up to two pixels (Fig. \ref{fig:degraded} c).
A narrower FOV can further improve the S/N ratio by temporal binning.

\subsection{Driving process of the jet}
\begin{figure}
 \includegraphics[width=\linewidth]{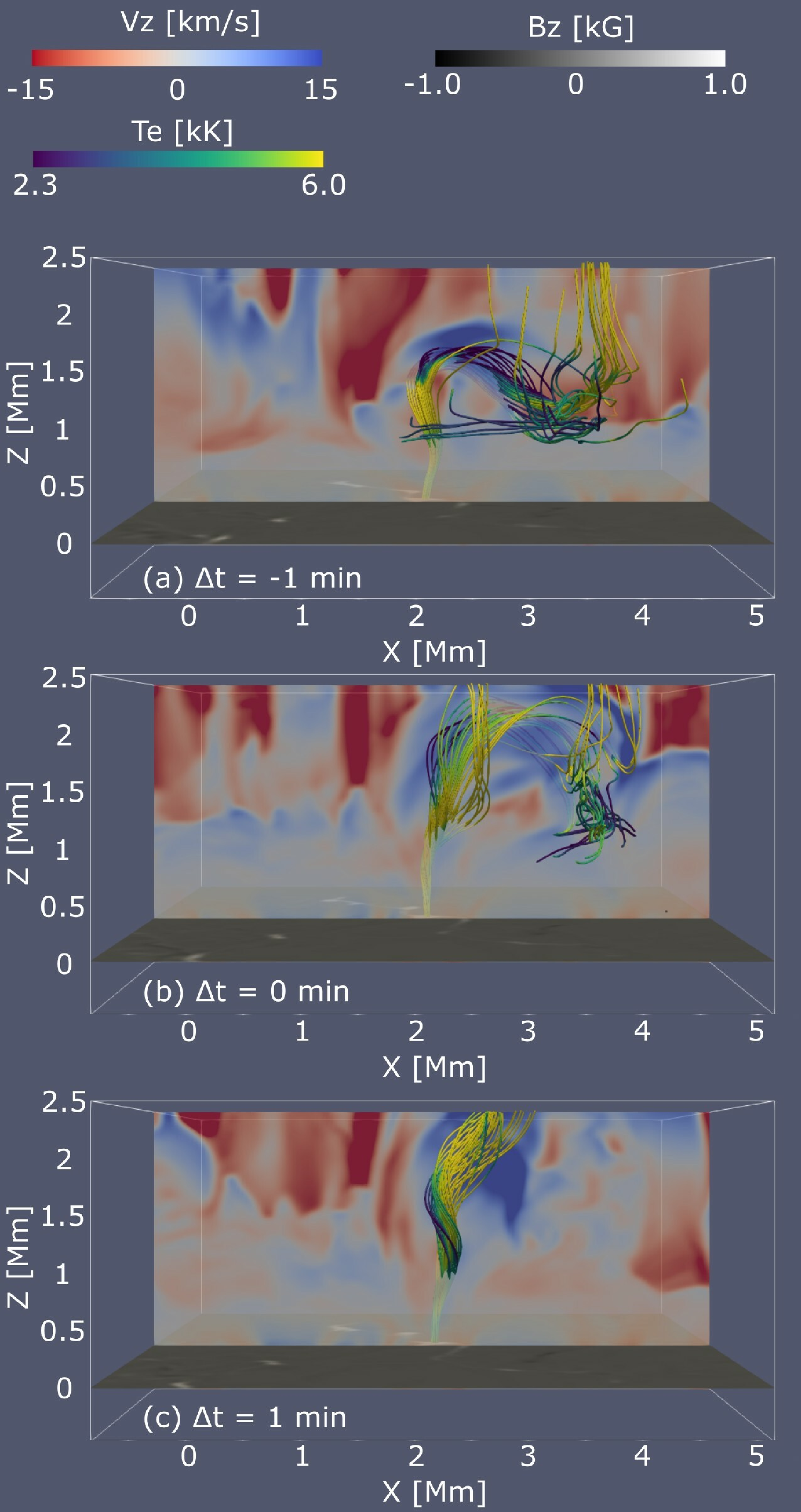}
 \caption{Evolution of the jet-related field lines. The magnetic field lines at $\Delta t$ = -1 min were temporally traced assuming ideal MHD conditions. The colours in x-y and x-z slices indicate $B_{\rm z}$ and $V_{\rm z}$, respectively. The colours on the field lines represent the local temperature.}
 \label{fig:jet_initiation}
\end{figure}

During the jet, the foot point of the twisted field sank and swirled counterclockwise from the edge to the centre of the magnetic concentration in the photosphere.
Simultaneously, the twisted part moved upward and became a vertical axial field (Fig. \ref{fig:jet_initiation}).
At $\Delta$t = -1 min, upflow A, which was about to collide with the transition region, was connected to the twisted field (Fig \ref{fig:jet_initiation}a).
We temporally traced a group of field lines that were connected to the upflowing material assuming an ideal MHD process, and found that the twisted field lines evolved into the axial field with time (Fig \ref{fig:jet_initiation}b \& c).
As the twisted part moved upward, part of the cool material inside the flux was lifted to the corona to form the jet, while the other half was flowing down along the axial field.
After the field lines became the flux tube's axial field, the ambient plasma continuously supplied another twisted field, which maintained the tornado-like structures, as observed in the horizontal field evolution (column g \& h in Fig. \ref{fig:xy_synthesis_map}).

There are several candidates for the triggering process of upflow A.
The first candidate is the flux merger at $\Delta$t = -4 min.
By temporally tracing field lines in sheets B \& C, we found that the field lines were swirling counterclockwise when approaching the intersection of 
the granular lanes.
Because B$_{\rm z}>0$ in the flux sheets, and the current density in the z-direction was negative (inferred from column g in Fig. \ref{fig:xy_synthesis_map}), 
the additional counterclockwise vorticity drove the upward propagating Alfv\'{e}nic motion.
This Alfv\'{e}nic motion produces the upflow A through nonlinear wave pressure \citep{1982SoPh...75...35H}.
The second candidate is the downflow along the axis of the flux tube.
This downflow contributes to the upflow via the rebounding process \citep{1988ApJ...327..950S,2016ApJ...827....7K}.
The third candidate is magnetic reconnections between the twisted field and its overlying field.
This reconnection may contribute to the upflow via sling shot effects, similar to the reconnection above the emerging flux \citep{1996PASJ...48..353Y}.
The combinations of the three candidates could contribute to the upflow, although the further theoretical investigation is needed to quantitatively measure the degree of contribution of each process.

\subsection{Summary of the evolving jet}
Herein, we briefly summarise the important physics and observables during the evolution of the chromospheric jet.
\begin{itemize}
    \item Before $\Delta$t = - 6 min, two flux sheets B and C were approaching the intersection of granular lanes. 
    This motion can be detectable by tracing enhanced MCP signals or bright I$_{\rm core}$ in \ion{Fe}{i} line (Fig. \ref{fig:xy_synthesis_map} \& \ref{fig:xy_synthesis_map2}).
    \item At $\Delta$t = - 4 min, the two flux sheets merged. 
    After the flux merger, the down-flowing hot core was continuously observed in the \ion{Fe}{i} line at the foot point during the lifetime of the jet 
    (Fig. \ref{fig:xy_synthesis_map2} \& \ref{fig:sit_and_stare}). 
    \item At $\Delta$t = 0 min, the twisted field line moved up to collide with the transition region (Fig. \ref{fig:mhd_evolution_2d} \& \ref{fig:jet_initiation}).
    The twisted field lines can be observed as the arc-like linear polarization signals with strong enough amplitude in \ion{Ca}{ii} line (Fig. \ref{fig:xy_synthesis_map} \& \ref{fig:degraded}).
    The arc-like structure was dark in the line core, showing blueshifted velocities (Fig. \ref{fig:sit_and_stare}). 
    Moreover, a bright core and a dark envelope in \ion{Ca}{ii} would be useful to specify the jet structure.
    \item The magnetic-tornado-like structure with a hot core and cool envelope continued to exist during the lifetime of the jet (0$<\Delta$t$<$9 min).
    This feature would be detectable from both I$_{\rm core}$, MLP, and TLP signals in \ion{Ca}{ii} (Fig. \ref{fig:xy_synthesis_map}, \ref{fig:xy_synthesis_map2}, \ref{fig:sit_and_stare}, and \ref{fig:degraded}).
\end{itemize}








\section{Conclusion }

Although there are several candidate mechanisms for generating chromospheric jets, it has been difficult to conclusively determine the driving mechanism directly from observations.
Because the magnetic field plays a key role, high spatial, spectral, and spectro-polarimetric observations are essential for distinguishing between competing mechanisms.
We found that flux sheet merger, bright and down-flowing core, and arc-like linear polarisation signals with dark and blue-shifted envelope were expected from chromospheric jets, although we analysed only one chromospheric jet appearing in the RMHD simulation.
These properties are the manifestation of the transition process from the ambient twisted field to the axial field, which is consistent with the upwardly propagating nonlinear Alfv\'{e}nic waves. 
The triggering process is still under investigation, although we considered that the key mechanisms would the flux merger,  downflow along the core, or magnetic reconnection above the twisted field.

Our prediction will be a useful tool for distinguishing the driving mechanisms of chromospheric jets when combined with future observations, such as SUNRISE III, DKIST, or EST. 
An appropriate time cadence, integration time, and field of view should be selected to fully capture the characteristic magnetic field associated with the chromospheric jets.

\section*{Acknowledgements}
We would like to express our sincere gratitude to the reviewer for the invaluable feedback and constructive suggestions on our manuscript.
Numerical computations for the radiation MHD simulation were performed on a Cray XC30 supercomputer at the Centre for Computational Astrophysics, National Astronomical Observatory of Japan. Numerical analyses and Stokes synthesis were performed on the analysis servers at the Centre for Computational Astrophysics at the National Astronomical Observatory of Japan. A part of this study was conducted using computational resources from the Centre for Integrated Data Science at the Institute for Space-Earth Environmental Research, Nagoya University, through a joint research program. 
This work was supported by JSPS KAKENHI Grant Number JP18H05234 (PI: Y. Katsukawa).
 
 \section*{Data Availability}
 The data underlying this article will be shared upon reasonable request by the corresponding author.





\bibliographystyle{mn2e}
\bibliography{mn2e-jour,myref}

\bsp	
\label{lastpage}
\end{document}